# Topological Metal MoP Nanowire for Interconnect

Hyeuk Jin Han[1,2], Sushant Kumar[3], Xiaoyang Ji[4], James L. Hart[1,2,5], Gangtae Jin[1,2], David J. Hynek[1,2], Quynh P. Sam[5], Vicky Hasse[6], Claudia Felser[6], David G. Cahill[4], Ravishankar Sundararaman[3], Judy J. Cha[1,2,5*]

[1]Department of Mechanical Engineering and Materials Science, Yale University, New Haven, Connecticut 06511, USA
[2]Energy Sciences Institute, Yale West Campus, West Haven, Connecticut 06516, USA
[3]Department of Materials Science and Engineering, Rensselaer Polytechnic Institute, Troy, New York 12180
[4]Departmnet of Materials Science and Engineering, University of Illinois Urbana-Champaign, Urbana, Illinois 61801, USA
[5]Department of Materials Science and Engineering, Cornell University, Ithaca, New York 14853, USA
[6]Max Planck Institute for Chemical Physics of Solids, Dresden, Germany

## Abstract

The increasing resistance of Cu interconnects for decreasing dimensions is a major challenge in continued downscaling of integrated circuits beyond the 7-nm technology node as it leads to unacceptable signal delays and power consumption in computing. The resistivity of Cu increases due to electron scattering at surfaces and grain boundaries of the interconnects at the nanoscale. Topological semimetals, owing to their topologically protected surface states and suppressed electron backscattering, are promising material candidates to potentially replace current Cu interconnects as low-resistance interconnects. Here, we report the attractive resistivity scaling of topological metal MoP nanowires and show that the resistivity values are comparable to those of Cu interconnects below 500 nm$^2$ cross-section areas. More importantly, we demonstrate that the dimensional scaling of MoP nanowires, in terms of line resistance versus total cross-sectional area, is superior to those of effective Cu and barrier-less Ru interconnects, suggesting MoP is an attractive solution to the current scaling challenge of Cu interconnects.

**Main Text:**

Interconnects are metal wires that connect transistors, transmit signals in computer chips, and occupy a large fraction of integrated circuits. In the early 2000s, copper (Cu) replaced aluminum as low-resistance interconnects for continued downscaling of integrated circuits.(*1, 2*) However, Cu can no longer support the dimensional reduction at the smallest feature size of interconnects due to its ever-increasing resistivity that stems from surface and grain boundary electron scattering.(*3*) The high resistivity of current Cu interconnects can account for up to 35% of total signal delays and nearly half of dynamic power dissipation in computer chips.(*4*) Thus, future energy-efficient computing technologies require breakthroughs in interconnect technologies,(*5*) particularly in new interconnect materials.

Topological semimetals are promising materials for low resistance interconnects as their topologically protected surface electrons are forbidden to backscatter.(*6–8*) Several experimental studies on nanostructured topological semimetals show promising results. The Weyl semimetal NbAs, for example, exhibits a factor of ten decrease in resistivity from bulk crystals (35 μΩ-cm) to nanobelts (~ 3 μΩ-cm) at room temperature.(*9*) Similarly, recent theoretical results from IBM predict that the multifold fermion system CoSi would exhibit lower resistivity than Cu at very small dimensions as the current conduction is dominated by Fermi-arc surface states.(*10*) Among recently reported topological metals, molybdenum phosphide (MoP) is a simple binary compound that was predicted (*11*) and experimentally confirmed to host topologically protected fermions.(*12*) Single crystal MoP shows extremely low resistivity and high carrier density,(*13*) representing an exciting opportunity to potentially replace Cu interconnects. In this work, we report systematic engineering and dimensional resistivity scaling of poly-crystalline MoP nanowires through the template assisted chemical vapor deposition (CVD) and demonstrate that MoP nanowires exhibit dimensional scaling superior to effective Cu (Cu with TaN liner) and Ru for dimensions beyond the 7-nm technology node.

**Results**

MoP has a WC-type hexagonal crystal structure (Figure 1A) with lattice parameters *a* = *b* = 3.22 Å and *c* = 3.18 Å.(*13*) Mo and P share the same coordination number and coordination environment of six in a trigonal prism. *Ab initio* calculations show the presence of topologically protected triple point fermions roughly 0.7 eV below the Fermi level along the Γ-A high symmetry **k**-point path in the Brillouin zone (Figure 1B), which were verified by previously published angle-resolved photoemission spectroscopy results.(*12*) Figure 1C shows the first-principles calculated Fermi surface of MoP along with the highly anisotropic electron-phonon mean free path distribution. While the droplet-shaped electron pockets (elongated along the Γ-A direction) are found to have the longest mean free paths, the flat hole pockets centered at Γ have the shortest mean free paths. Based on the calculated Fermi surface, we compute the electron mean free path length of 10.5 nm and the room temperature resistivity of 12.9 μΩ·cm along the *a*-axis and 9.8 μΩ·cm along the *c*-axis for bulk MoP. The mean free path of MoP is nearly four times shorter than that of Cu (40 nm), which is advantageous for interconnect applications because the dimensional increase of resistivity will not emerge until the interconnect dimension approaches the mean free path.

Considering electron scattering at surfaces of a square cross-section wire, we calculate the resistivity increase of MoP nanowires with decreasing wire width and compare it to that of Cu using the Fuchs-Sondhemier models (*14*, *15*) (Figure 1D). This calculation assumes single-crystalline Cu and MoP, and thus only includes electron scattering at surfaces of the wire and excludes scattering at grain boundaries. The different curves for MoP ($p$ = 0, 0.5, 1) correspond to the different values of the specularity parameter $p$, which is the proportion of electrons scattered elastically at the surfaces. $p$ = 1 corresponds to a perfectly smooth surface with specular reflection of electrons, while $p$ = 0 represents completely diffuse scattering. For Cu, we used reported values for surface scattering (*16*, *17*) and this case represents the idealized case without grain boundary scattering.(*18*) Our calculations predict that MoP wires will outperform Cu wires with liner below 9 nm widths for ideal characteristics ($p$ = 1) and below 7 nm for worst-case characteristics ($p$ = 0) (more details can be found in Supporting Information). The latest 5-nm technology node contains Cu interconnects with ~ 15 nm width in the M0 stack;(*19*, *20*) thus, our calculations suggest MoP as a promising interconnect material for future technology nodes.

We synthesize MoP nanowires by heating $MoO_3$ nanowires in the presence of $PH_3$ vapors and $H_2$ gas in a CVD system (see Supplementary Figure S1 and Methods for details). Transmission electron microscopy (TEM) images of MoP nanowires show poly-crystalline wires (Figure 2A). Figure 2B shows an atomic-resolution high-angle annular dark-field scanning TEM (HAADF-

STEM) image of a MoP grain projected along the [010] direction, with a lattice spacing of 3.2 Å in agreement with the MoP (001) lattice spacing. Selective area electron diffraction (SAED) patterns and X-ray diffraction data (Supplementary Figure S2) confirm the atomic structure of MoP for the nanowires. Energy dispersive spectroscopy (EDS) analysis of the poly-crystalline nanowires indicates that the Mo:P atomic ratio is close to 1:1 as the EDS data from the nanowires overlap with that from a MoP bulk crystal grown via chemical vapor transport (CVT) (Figure 2C). Figure 2D shows Raman spectra taken from the MoP nanowires and bulk crystal; one main peak is resolved at ~ 407 cm$^{-1}$, corresponding to the E mode.(*21*, *22*)

We focus our attention on room temperature transport properties of MoP nanowires to test the feasibility of MoP as a low-resistance interconnect material. Nanodevices were fabricated using Cr/Au contacts to measure the resistance of MoP nanowires as a function of wire diameter. Figure 3A shows two-probe *I–V* curves of these MoP nanowires, which are linear and ohmic as expected for a metal. Four-probe resistance measurements were carried out to remove the contact resistance and the measured resistance was converted to resistivity where the cross-section areas of the MoP nanowires were obtained by measuring the wire diameters using scanning electron microscopy and assuming a circular cross-section. Figure 3B summarizes the resistivity of MoP nanowires at room temperature. Remarkably, the resistivity of MoP nanowires ranges between 11 and 13 μΩ·cm for cross-section areas of 300 - 1500 nm$^2$, which is in agreement with the calculated resistivity and represents only ~ 50% increase from the bulk resistivity of single crystal MoP (8 μΩ·cm).(*13*) By contrast, the resistivity of Cu interconnects with a liner at 500 nm$^2$ cross-section area increases by more than seven times from the bulk value (~12 μΩ·cm from 1.68 μΩ·cm).(*3*) Thus, we show that the resistivity of MoP nanowires is already as good as that of Cu interconnects in these nanoscale dimensions.

The resistivity of poly-crystalline MoP nanowires is high for large wires (~30 μΩ·cm for 3000 nm$^2$ cross-section area in Figure 3B) and decreases asymptotically to the bulk resistivity value as the cross-section area decreases. This observation is attributed to the high number of grain boundaries present in large MoP nanowires, many of which are parallel to the direction of current flow and dramatically increase grain boundary electron scattering, as illustrated in Figure 3C. When the grain boundaries parallel to the current direction are absent, for example in small MoP wires (Figure 2A), the resistivity of MoP nanowires is close to the bulk resistivity. The diameter-dependent crystalline quality of MoP nanowires is supported by analyzing the residual resistance ratio (RRR). Figure 3D shows the temperature dependent resistivity curves of MoP nanowires with varying diameter. The RRR decreases with increasing wire diameter (inset in Figure 3D), which supports the observed resistivity trend and agrees with the microstructure difference between large and small MoP nanowires. The microstructure difference was directly verified using TEM, where we observe a sudden increase in the number of grain boundaries at the nanowire width of 45 nm (Figure 3E), which coincides with the nanowire width at which the resistivity of MoP nanowires starts to decrease rapidly. Direct correlation between the microstructure and transport properties for two MoP nanowires are shown in Supplementary Figure S3.

In Figure 4A, we benchmark the resistivity of our best performing poly-crystalline MoP nanowires against current state-of-art interconnect materials, such as Cu with improved liner (Mn/Ru) and barrier-less Ru for cross-section areas < 1000 nm$^2$,(*3, 23*) which represents dimensions below the 7-nm back end of line (BEOL) technology nodes. Our poly-crystalline MoP nanowires show the lowest resistivity values with the best dimensional scaling behavior. The room temperature resistivity measurements thus experimentally demonstrate that electron scattering at surfaces and grain boundaries of MoP nanowires are negligible at these dimensions, in contrast to Cu, owing to the much smaller electron mean free path of MoP (10.5 nm) as compared to Cu (40 nm). The resistivity of MoP nanowires is also lower than that of Ru, the leading material candidate to replace Cu interconnects. Given the polycrystalline nature of our MoP nanowires, we repeat the calculation of the resistivity increase of the MoP and Cu nanowires with decreasing wire width and include grain boundary scattering, which shows that MoP will still outperform Cu wires with liner below 12 nm widths (Supplementary Figure S4). We also compared temperature coefficient of resistivity values of MoP nanowires with those of Cu and Ru and determined that electron scattering at surfaces and grain boundaries is not as detrimental for MoP as for Cu and Ru (Supplementary Figure S5). Thus, this benchmark comparison suggests that MoP is superior to Cu and Ru for cross-section areas < 300 nm$^2$.(*3, 23*)

The superior dimensional scaling of MoP resistivity over effective Cu and Ru is made apparent by analyzing the line resistance. For interconnect applications, the rate of increase of line resistance needs to be as small as possible as the interconnect dimensions shrink. Figure 4B shows the line resistance of MoP, effective Cu, and Ru as a function of the wire cross-section area, plotted in logarithmic scale.(*3*) The difference in line resistance increase with decreasing wire cross-section for MoP versus Ru and Cu is clear. The dotted lines are fit to the experimental data, which show log(R) scales linearly with log (cross-section area). The fitted slope shows that the increase is most severe for Cu and least severe for MoP. Thus, we anticipate that MoP will outperform Cu as low-resistance interconnects below the 7-nm technology node.

If MoP were to replace Cu in low-resistance interconnects, other materials properties must also be considered, such as surface oxidation, thermal conductivity, and electromigration. If MoP oxidizes easily or electromigrates under the application of electrical field, MoP would require a barrier layer, which is often resistive and would negate the observed attractive properties of MoP nanowires. Moreover, if the thermal conductivity of MoP is low, then effective heat management would be difficult.

Using time-domain thermoreflectance (TDTR), an averaged thermal conductivity of 99 W/m-K with a standard deviation of 2 W/m-K was obtained from measuring several MoP nanoplates at 300 K (Supplementary Figure S6 and Table S1); the systematic uncertainty of TDTR measurements on the through-plane thermal conductivity is usually 7%.(*24*) The TDTR measurement of the nanoplates is done on the (001) growth facet and is therefore mostly sensitive to the thermal conductivity along the <001> direction of the crystal. The measured thermal conductivity of MoP nanoplates is in good agreement with that of a CVT-grown MoP bulk crystal, which was measured to be 96 W/m-K using TDTR. The TDTR measurement of the bulk crystal was done on a growth facet of the crystal at 300 K; we did not determine the orientation of this facet and therefore the thermal conductivity we measured for the bulk crystal is for an unknown direction of the crystal. (Supplementary Figure S6). These thermal conductivity values are in line with calculations, but an order of magnitude lower than the thermal conductivity reported by Kumar *et. al.* using CVT-grown MoP bulk crystals.(*13*, *25*)

 For surface oxidation, we annealed the MoP nanowires in air up to 400 °C and observed negligible surface oxidation up to 150 °C based on X-ray photoelectron spectroscopy (Supplementary Figure

S7). The negligible surface oxidation for MoP nanowires was confirmed by measuring the resistance of MoP nanowire devices left in ambient. The resistance did not increase for the three MoP nanowires we tested (diameters of 37 nm, 43 nm, and 58 nm) over the 48 hrs they were left in ambient, while the resistance increased by 140 % for a 20 nm-thick Cu film under the same condition (Supplementary Figure S7).

Is there another topological semimetal that may be better than MoP for interconnect applications? To answer this, we surveyed reported values of room temperature transport properties of topological semimetals. Figure 4C summarizes the carrier density versus resistivity of topological semimetals from bulk and sub-micron samples. For interconnect applications, high carrier density and low resistivity are desirable. For this reason, topological insulators are not suitable as interconnects as they have high resistivity values. We observe that MoP has the best attributes for interconnect applications among the topological semimetals reported in literature. We note, however, that it remains to be seen if another topological semimetal may outperform MoP at small dimensions due to more prominent contributions from their topological surface states.

The room temperature transport data of MoP nanowires do not suggest any obvious effects from the topological surface states or suppression of electron backscattering (see Supplementary Figure S8 and S9 for additional calculations of MoP thin slabs and discussions on the lack of topological effects). Regardless, the resistivity values of our MoP nanowires are already lower than those of Cu interconnects below 500 $nm^2$ cross-section areas, and the resistivity scaling behavior of MoP nanowires is superior to those of Cu and Ru interconnects. Thus, our work demonstrates MoP as a breakthrough material for interconnect technologies for continued downscaling of integrated circuits and future energy-efficient computing.

## Experimental

### Synthesis of MoP nanowires

We use $MoO_3$ nanowires as the precursor to synthesize MoP nanowires. $MoO_3$ nanowires were grown by CVD, as previously reported by our group.(26) 0.15 g of the $MoO_3$ source powder (Sigma-Aldrich, 99.95%) was placed at the center of a 1 in. tube furnace with anodized aluminum oxide (AAO, InRedox) substrates located 14 cm downstream. After purging with Ar, the system was pumped down to 200 mTorr, and then $H_2$ was flowed at 20 sccm, bringing the furnace pressure to 5 Torr. The furnace was heated to 600 °C in 15 min and held at that temperature for 10 min to produce $MoO_3$ nanowires with high yield. MoP nanowires were synthesized by converting $MoO_3$ nanowires. $MoO_3$ nanowires were placed in a tube furnace with a sufficient amount (3 g) of $NaH_2PO_2 \cdot H_2O$ (Sigma-Aldrich, ≥99%) placed upstream (15 - 17 cm from the center of the furnace). After purging with Ar, the system was pumped down to 200 mTorr, and then $H_2$ was flowed at 20 sccm, bringing the furnace pressure to atmospheric pressure. The furnace was heated to 700 °C in 30 min, held there for 1 hr, and then cooled down to room temperature naturally.

### Characterization of MoP nanowires

Structural characterization of the MoP nanowires was carried out using SEM and TEM. A field emission SEM (Hitachi S-4800) was used at an acceleration voltage of 10 kV and a working distance of 5 mm. High-resolution TEM images were obtained using a 200 kV accelerating voltage TEM (FEI, Talos F200X). Atomic-resolution STEM images were obtained using a probe-corrected microscope (ThermoFisher Scientific, Spectra 300) at 200 kV. Raman spectra were obtained using a Horiba LabRAM HR Evolution Spectrometer with an excitation wavelength of 532 nm. For chemical compositions of the nanowires, X-ray photoelectron spectroscopy (XPS) data were acquired using a multipurpose X-ray photoelectron spectrometer (Sigma Probe; Thermo VG Scientific). The X-ray diffraction (XRD) measurements were carried out using a multipurpose thin-film X-ray diffractometer (D/Max 2500; Rigaku).

### Device fabrication and transport measurements

Synthesized MoP nanowires were transferred onto $SiO_2$/Si substrates by stamping and coated with triple e-beam resist layers (~200 nm PMMA A3 as the first layer, ~200 nm MMA EL 8.5 as the second layer, and ~200 nm PMMA A3 as the third layer). Electrode patterns were obtained by standard e-beam lithography using a Vistec EBPG 5000+. The devices were designed for four-probe measurements, and the distance between the electrodes was kept at ~200 nm. After the pattern was developed, 10/100 nm-thick Cr/Au electrical contacts were deposited by e-beam evaporation. Transport measurements at low temperature were performed using a Quantum Design Dynacool physical property measurement system equipped with a base temperature of 2 K. Transport data were taken at applied currents ranging between 10 µA and 100 µA.

**CVD growth of 2D MoP single crystal nanoplates for thermal conductivity measurements**

A liquid droplet of gallium (Ga, Sigma-Aldrich, 99.9995%) was placed onto 1 cm×1 cm molybdenum (Mo) foil (Sigma-Aldrich, 99.9%) substrate. The Ga–Mo substrate was then heated in a quartz tube to 1100 °C at a heating rate of 30 °C min$^{-1}$. At 1100 °C, red phosphorus powder (Sigma-Aldrich, ≥99.99%) was introduced into the tube furnace downstream where the temperature was around 400 °C. The furnace was kept at 1100 °C for 20 min under the flow of Ar (250 sccm) and $H_2$ (50 sccm).

**Theoretical calculations**

We used open-source planewave Density Functional Theory (DFT) code JDFTx (*27*) to compute the bulk resistivity and resistivity scaling of MoP and Cu. The electronic structure of MoP was calculated using the fully relativistic revised Perdew-Burke-Ernzerhof (PBEsol) (*28*) pseudopotentials and generalized gradient approximation (GGA) exchange-correlational functional. We used a planewave cut-off of 35 Hartrees and a Fermi smearing of 0.01 Hartrees. A **k**-mesh of $12 \times 12 \times 12$ and a **q**-mesh of $3 \times 3 \times 3$ were employed for the electron and phonon calculations respectively. Subsequently, the electronic states, phonon modes, and electron-phonon matrix elements are transformed to the maximally localized Wannier function basis (*29*) and interpolated to a very fine mesh (~$10^5$ points) to obtain converged integrals for the linearized Boltzmann solution for bulk resistivity. The electron-phonon lifetimes are computed using the Fermi's golden rule. For details of the implementation of these methods, refer to (*30*) and (*31*). The resistivity scaling for single-crystalline square wires was calculated using an asymptotic expansion of the Boltzmann solution as detailed in (*32*). Additional calculations for thin slabs of MoP are shown in Supplementary Figure S8 and S9.

**Thermal conductivity measurements**

The thermal conductivities of MoP nanoplates and a bulk crystal were determined by TDTR measurements. An Al layer (~80-100 nm) was coated on the surface of the MoP sample as the transducer layer. A mode-locked Ti:Sapphire laser with a wavelength of ~785 nm was split into the pump beam and the probe beam by optical filters. The pump beam was modulated by an electro-optical modulator with a frequency of 9.3 MHz, generating a temperature rise on the sample surface. The probe beam was modulated with a frequency of 200 Hz and to detect the temperature change on the sample surface after a delay time. The pump and probe beam were focused on the sample surface by an objective lens (20× for MoP nanoplates and 10× for MoP bulk). The reflectance of the probe beam was picked up at the frequency of the pump beam by an RF lock-in amplifier, then two computer-based lock-ins detected the RF lock-in outputs at the frequency of the probe beam. The ratio signals ($-V_{in}/V_{out}$, $V_{in}$: in-phase voltage signal, $V_{out}$: out-of-phase voltage signal) were collected and then fit to an analytical solution of a heat transfer equation for a multilayer model of the sample structure.

In data analysis, the structure of MoP nanoplates was Al/MoP/SiO$_2$/Si, and the thermal conductance of the Al/MoP interface and the thermal conductivity of MoP were fit as unknown parameters. The thermal conductivity of Al was 166 W/m-K, measured by a four-point method and calculated from the Wiedemann-Franz law. The volumetric heat capacities of Al and MoP were from previous literatures, which are 2.43 and 2.50 J/cm$^3$-K, respectively.(*13*, *33*) The thermal conductance of the MoP/SiO$_2$ interface was set to be 10 MW/m$^2$-K, although the MoP nanoplate was thermally thick and the TDTR measurements were not sensitive to the segments below MoP. As for the MoP bulk sample, the structure was Al/MoP, and the thermal conductivity of Al was 162 W/m-K.

**Acknowledgments:** H. J. H. and S. K. gratefully acknowledge the support by the Semiconductor Research Corporation (SRC) – nCORE IMPACT center. J. C. C. acknowledges support from Betty & Gordon Moore Foundation under the EPiQS Investigator Award. Theoretical calculations were performed at the Center for Computational Innovations (CCI) at Rensselaer Polytechnic Institute. X.J. thanks Sushant Mahat for thickness measurements of MoP nanoplates.

**Author information**

**Corresponding Authors**

*E-mail: judy.cha@cornell.edu

**Author contributions:** H. J. Han and J. J. Cha conceived the project and designed experiments. H. J. Han optimized the fabrication, prepared samples, and carried out structural and chemical characterizations. S. Kumar and R. Sundararaman carried out *ab initio* calculations and computed the dimensional scaling of the resistivity of MoP and Cu wires. H. J. Han, J. L. Hart, and Q. P. Sam carried out the TEM characterization. G. Jin and D. J. Hynek carried out further materials characterization on MoP nanowires. X. Ji and D. G. Cahill measured the thermal conductivity of MoP plates. V. Hasse and C. Felser synthesized the bulk crystals of MoP. H. J. Han performed the electrical properties experiments and analyzed the data. H. J. Han and J. J. Cha led the discussions and wrote the manuscript with input from all authors. J. J. Cha supervised the project.

**Competing interests:** The authors declare no competing financial interests.

**Data and materials availability:** Data supporting the plots and other findings in this Article are available from the authors upon reasonable request.

**Supplementary Materials**

Materials and Methods

Figs. S1 to S9

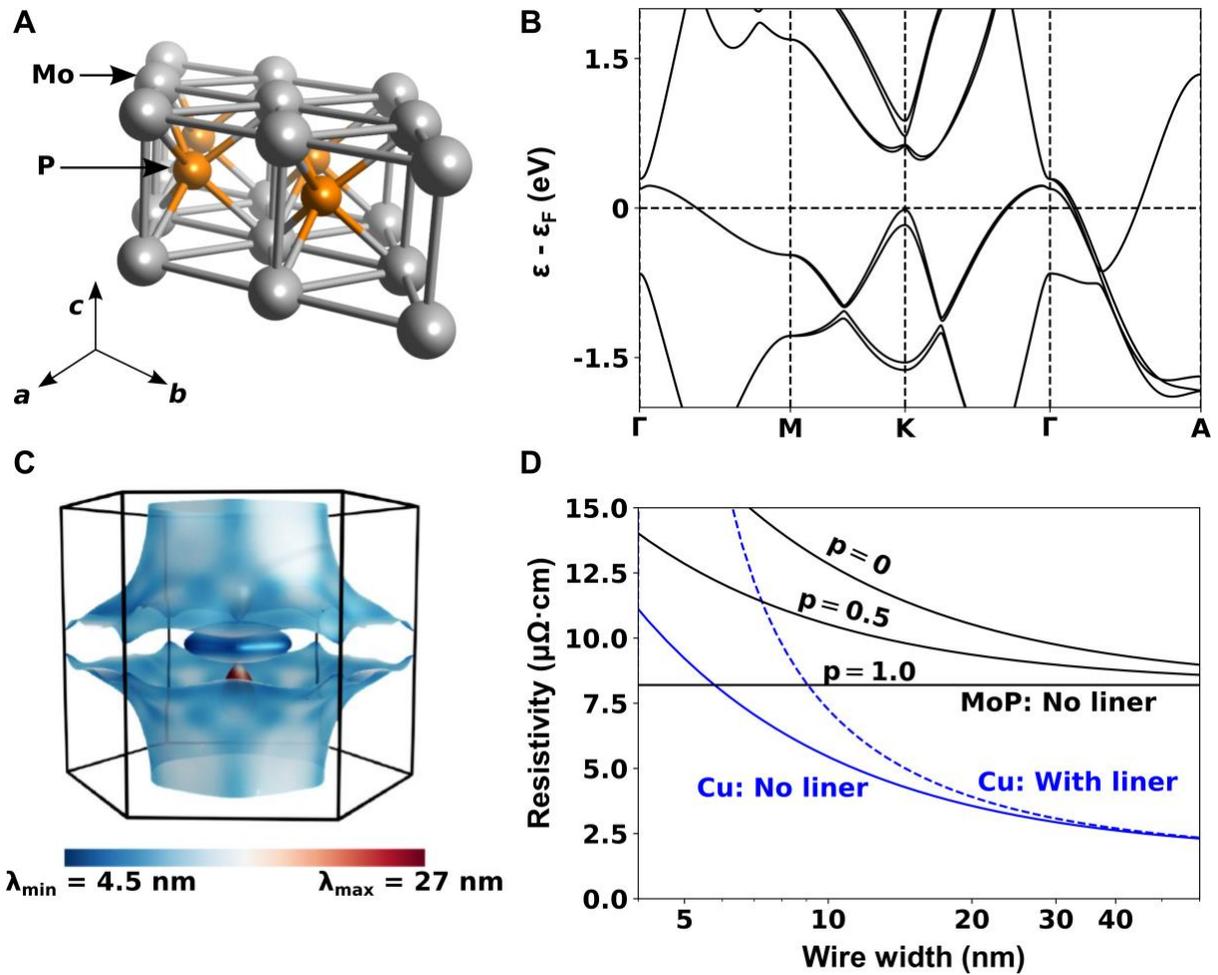

**Figure 1: Crystal structure, electronic band structure, Fermi surface and calculated resistivity of MoP.** (A) A schematic of the hexagonal crystal structure of MoP. (B) Calculated electronic band structures along high-symmetry **k**-point paths. (C) Fermi surface of MoP with calculated electron mean free path lengths. The polyhedron represents the Brillouin zone. (D) Calculated resistivity scaling of MoP and Cu wires. Here, experimentally measured bulk resistivities $r_0$ ($r$ when width $\rightarrow \infty$) for Cu and MoP are used. The curves for Cu correspond to p = 0.

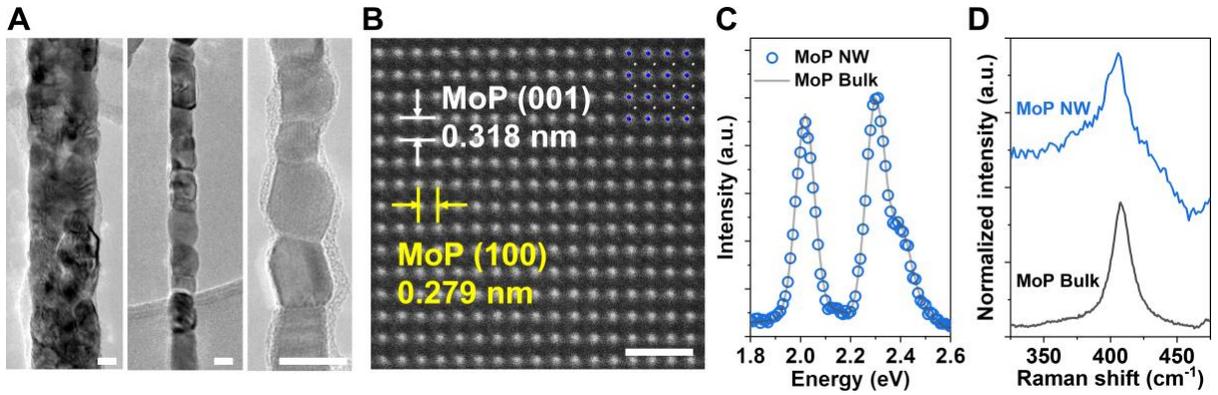

**Figure 2: Characterization of MoP nanowires.** (A) TEM images of MoP nanowires of varying diameters. Scale bars, 20 nm. (B) Atomic-resolution STEM image of a MoP nanowire showing high crystalline quality in single grains observed along the [010] direction, with bright Mo atomic columns. Scale bar, 1 nm. (inset) Atomic structure model of MoP viewed along the [010] direction (Blue: Mo atoms, White: P atoms). (C) TEM-EDS spectra acquired from a MoP nanowire and reference MoP bulk crystal. (D) Raman spectra acquired from MoP nanowires and reference MoP bulk; one phonon mode is identified. a.u., arbitrary units.

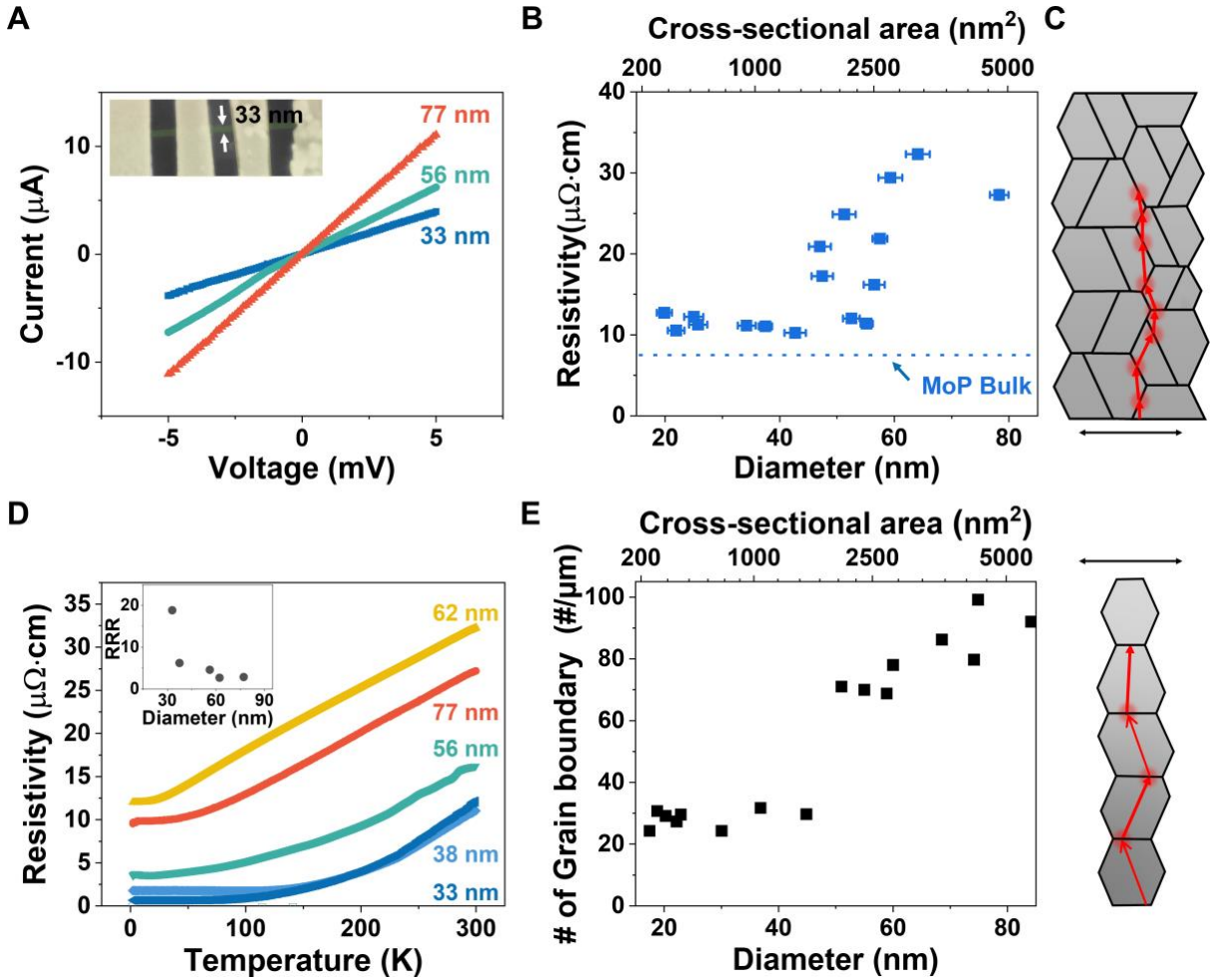

**Figure 3: Electron transport properties of MoP nanowires.** (A) *I–V* curves of poly-crystalline MoP nanowires with decreasing diameter. Inset: SEM image of a 33 nm-diameter MoP nanowire device. (B) Room temperature resistivity data of MoP nanowires with varying diameter (cross-sectional area). (C) Schematics of MoP nanowires with diameter > 45 nm (top) and < 45 nm (bottom), respectively. For MoP nanowires with large diameters, grain boundaries that are parallel to current flow are present, increasing the resistivity. The horizontal arrows denote 45 nm. (D) Temperature-resistivity (ρ) curves of MoP nanowires with varying diameter. Inset: Residual resistance ratio (RRR) versus nanowire diameter. (E) Number of grain boundaries per unit length present as a function of the diameter of MoP nanowires.

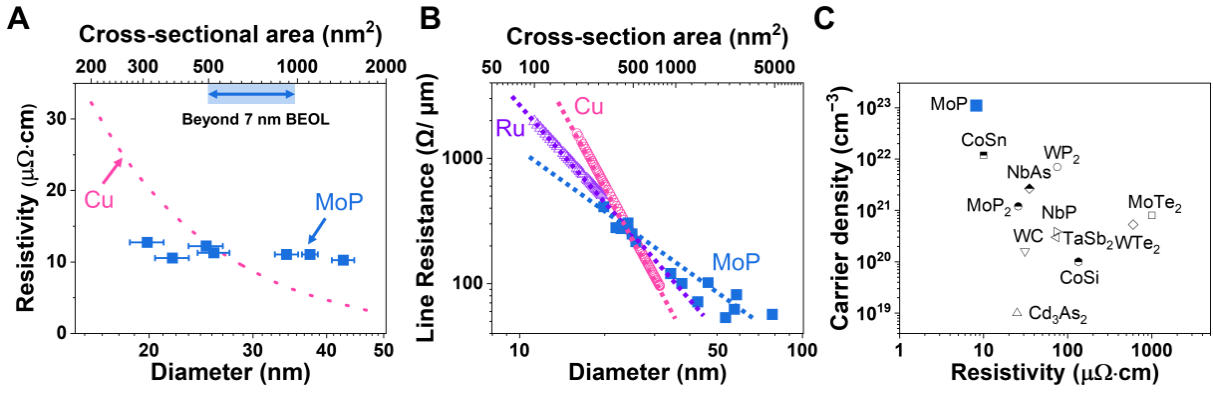

**Figure 4: Benchmark and prospect of MoP as low-resistance interconnects.** (A) Experimental resistivity data of MoP nanowires (blue square) and Cu wires (dotted line) below 7 nm BEOL nodes. (B) Line resistance of the Ru, Cu and MoP nanowires as a function of diameter (cross-section area). Data for Cu and Ru were taken from (*3*, *23*). (C) Comparison chart of carrier density and resistivity of topological semimetals from literature.